\documentclass[12pt]{article}
\usepackage{amsmath}
\usepackage{amsfonts}
\usepackage{amssymb}
\usepackage{graphicx}

\begin{document}

\pagestyle{myheadings}
\parindent 0mm
\parskip 6pt

\title{The O$(n)$ model on the annulus}

\author{John Cardy\\
Rudolf Peierls Centre for Theoretical Physics\\
         1 Keble Road, Oxford OX1 3NP,
         U.K.\footnote{Address for correspondence.}\\
and All Souls College, Oxford.}
\date{April 2006}
\maketitle
\begin{abstract}
We use Coulomb gas methods to derive an explicit form for the
scaling limit of the partition function of the critical O$(n)$
model on an annulus, with free boundary conditions, as a function
of its modulus. This correctly takes into account the magnetic
charge asymmetry and the decoupling of the null states. It agrees
with an earlier conjecture based on Bethe ansatz and quantum group
symmetry, and with all known results for special values of $n$. It
gives new formulae for percolation (the probability that a cluster
connects the two opposite boundaries) and for self-avoiding loops
(the partition function for a single loop wrapping non-trivially
around the annulus.) The limit $n\to0$ also gives explicit
examples of partition functions in logarithmic conformal field
theory.
\end{abstract}
\newpage
\section{Introduction.}
\label{sec1} The Coulomb gas approach to two-dimensional critical
models which can be written as loop gases, such as the O$(n)$
model and $Q$-state Potts models, has been extraordinarily
successful in deriving their universal bulk properties. First
developed by den Nijs \cite{denNijs} and Nienhuis \cite{Nienhuis}
in order to explain conjectured exact values for the principal
bulk critical exponents, it was adapted by di Francesco, Saleur
and Zuber \cite{diFSZ} to compute the partition function on the
torus, which encodes all the bulk scaling dimensions \cite{JC86}.

However, this particular approach has not, so far, been
successfully adapted to explain the conjectured exact values
\cite{JC84,DupSal} for the \em boundary \em scaling dimensions in
these models, even though many of them have now been derived using
other methods \cite{SalBau} applied to related lattice models,
and, more recently, rigorously using Schramm-Loewner Evolution
(SLE) \cite{SLE}. From the point of view of conformal field theory
(CFT), both the boundary and the bulk exponents are encoded in the
partition function on the annulus \cite{JC89}, and therefore a
direct computation of this object is of great interest. From the
point of view of CFT, general values of $n$ and $Q$ give
irrational non-unitary examples.

As we shall explain, the naive application of Coulomb gas methods
to domains with boundaries fails to account either for the
reduction of the central charge $c$ from its free field value of
unity, or for the elimination of the null states, which, as is
known from CFT, is necessary to maintain modular invariance.  In
this paper we propose a resolution of these issues which provides
an explicit formula for the partition function on the annulus.

Consider an annulus $(0\leq x<\ell,0<y<L)$, identifying
$x=0,\ell$. Introduce the conjugate moduli
$$
q=e^{-\pi\ell/L},\qquad \tilde q=e^{-2\pi L/\ell}\,.
$$
Note that if the annulus is conformally mapped to the region
$R_1<r<R_2$ between two circles, $\tilde q=\ln(R_2/R_1)$. We
impose free boundary conditions on the O$(n)$ spins on $y=0,L$. As
usual in the Coulomb gas, we introduce the parametrisation
$n=\sqrt Q=2\cos\chi$, $g=1-\chi/\pi$, where $1\leq g\leq 2$
corresponds to the dilute critical point of the O$(n)$ model (or
the tricritical point of the $Q$-state Potts model), and
$\frac12\leq g<1$ to the critical dense phase of the O$(n)$ model
(or the ordinary critical point of the Potts model.) Then our main
result for the annulus partition function is
\begin{equation}
\label{2} Z=q^{-\frac c{24}}\prod_{r=1}^\infty(1-q^r)^{-1}
\sum_{p\in\mathbb Z}\frac{\sin(p+1)\chi}{\sin\chi} \, q^{\frac
{gp^2}4-\frac{(1-g)p}2}\,.
\end{equation}

In terms of the conjugate modulus $\tilde q$ this becomes
\begin{equation}
\label{mod}
 Z=(2/g)^{1/2}\,{\tilde q}^{-\frac
c{12}}\prod_{r=1}^\infty(1-{\tilde q}^{2r})^{-1} \sum_{m\in\mathbb
Z}\frac{\sin((\chi+2m\pi)/g)}{\sin\chi}\, {\tilde
q}^{\frac{(\chi+2\pi m)^2}{2\pi^2g}-\frac{(1-g)^2}{2g}}\,.
\end{equation}

This has the form expected from boundary conformal field theory
\cite{JC89}: (\ref{2}) is\\
$Z=q^{-c/24}\sum_{\Delta}d_{\Delta}\chi_{\Delta}(q)$, where
${\Delta}$ runs over the allowed set of boundary scaling
dimensions, $\chi_{\Delta}(q)$ is a highest weight Virasoro
character, and $d_{\Delta}$ is a degeneracy factor, which is a
polynomial in $n$ (although only integer in those cases when the
theory is unitary). In this form the explicit expression (\ref{2})
is not new, and indeed is originally due to Saleur and Bauer
\cite{SalBau}, who deduced the allowed scaling dimensions from
Bethe ansatz and quantum group arguments. In this context the
degeneracy factor $d_\Delta$ is the quantum dimension. It has been
used extensively in papers by Saleur and Pasquier \cite{SalPasq},
Saleur \cite{Sal}, and more recently appeared in a paper by Read
and Saleur \cite{ReadSal}. However, a direct derivation from the
lattice O$(n)$ model solely using Coulomb gas arguments has not
appeared to our knowledge, and this is the point of the present
paper.

Note that (\ref{mod})  also has the form expected from boundary
CFT \cite{JC89}, namely\\ ${\tilde
q}^{-c/12}\sum_{\Delta}|b_{\Delta}|^2\chi_{\Delta}({\tilde q}^2)$
where now the sum is over allowed bulk scaling dimensions
$2{\Delta}$, and $b_{\Delta}$ is a matrix element with a boundary
state. In particular, for ${\Delta}=0$ we have
\begin{equation}
\label{gfn} b_0^2=-(2/g)^{1/2}\frac{\sin(\pi/g)}{\sin\pi g}\,,
\end{equation}
which gives the boundary entropy \cite{AffLud} $\ln b_0$.

We have explicitly written the dependence on $\chi$ in
Eqs.~(\ref{2},\ref{mod}) because, if we wish to consider a
modified partition function in which the loops which wrap
non-trivially around the annulus are counted with a different
weight $n'=2\cos\chi'$, it is simply necessary to replace
$\chi\to\chi'$. This allows us to compute interesting quantities
for percolation and self-avoiding loops.

For example, in critical percolation ($Q=1$) the probability that
a cluster connects the two boundaries of the annulus is
\begin{equation}
\label{crossing} P=\prod_{r=1}^\infty(1-q^r)^{-1}\sum_{k\in\mathbb
Z}(q^{\frac{8k^2}3-\frac{2k}3} -q^{\frac{8k^2}3+2k+\frac13})\,.
\end{equation}

The partition function for a single self-avoiding loop which wraps
non-trivially around the annulus (the number of such loops
weighted by $\mu^{-{\rm length}}$, where $\mu$ is the
non-universal connective constant) is
\begin{equation}
\label{4} Z_1=\prod_{r=1}^\infty(1-q^r)^{-1}\sum_{k\in\mathbb
Z}k(-1)^{k-1}\,q^{\frac{3k^2}2-k+\frac18}\,.
\end{equation}
In the limit $\tilde q\to0$, we find $Z_1\sim(1/6\pi)|\ln\tilde
q|$. The form of this agrees with a rigorous result of Werner
\cite{Werner05}.

The layout of this paper is as follows. In Sec.~\ref{sec2} we give
a brief survey of Coulomb gas methods as applied in the plane and
the cylinder, and then discuss the particular problems associated
with domains with boundaries, in particular the annulus. This will
lead to the proposal (\ref{2}) for the partition function. As with
most Coulomb gas methods, this is not wholly deductive, but relies
on some heuristic reasoning.  However, in Sec.\ref{sec3}, we show
that (\ref{2}) agrees with previously known results for various
special cases (for example the Ising model and the 3-state Potts
model.) In Sec.~\ref{sec4} we then derive a variety of new
results, some of which have already been mentioned above.

\section{Coulomb gas on the cylinder and the annulus.}
\label{sec2} \subsection{Basics.} We first summarise the Coulomb
gas arguments as applied to the plane and cylinder, as formulated
by de Nijs \cite{denNijs} and Nienhuis \cite{Nienhuis}, and
elaborated by Kondev \cite{Kondev}.

The O$(n)$ model is most easily realised on the honeycomb lattice.
At each site $r$ is an $n$-component spin ${\bf s}(r)$ (initially
$n$ is a positive integer.) The Boltzmann weight for a given
configuration is
\begin{equation}
\label{On} \prod_{r,r'}\big(1+t\,{\bf s}(r)\cdot{\bf
s}(r')\big)\,,
\end{equation}
where the product is over all edges $(r,r')$ of the lattice. The
partition function is the trace over these weights, a linear
operation defined by ${\rm Tr}\,1=1$, ${\rm
Tr}\,s_a(r)s_b(r)=\delta_{ab}$ and ${\rm Tr}\,s_a(r)={\rm
Tr}\,s_a(r)s_b(r)s_c(r)=0$. Expanding (\ref{On}) in powers of $t$
gives a sum over all subsets $\cal G$ of the edges, with an
associated factor $t^{|{\cal G}|}$. Implementing the trace
operation eliminates all subgraphs which are not unions of
non-intersecting closed loops (for the time being we ignore
boundaries), and each of these gets counted with a weight $n$.

At this point we can allow $n$ to be any positive real number.
This gives a measure on the allowed subgraphs $\cal G$, called the
loop gas. If $t$ is small, the mean loop length is finite, even in
the thermodynamic limit, but there is a critical value $t_c$ at
which it first diverges. This is called the dilute critical point.
For $t>t_c$ a single loop contains a finite fraction of the sites:
this is the dense phase.

The critical $Q$-state Potts model on the square lattice can also
be written, via the Fortuin-Kasteleyn \cite{FK} correspondence, in
terms of a loop gas, in which each closed loop carries a factor
\cite{Nienhuis} $\sqrt Q$.

Both these loop gas models can be mapped to a model of heights
$h(R)$ on the sites $R$ of the dual lattice, by first orienting
each loop, so that a configuration of $N$ non-oriented loops
corresponds to $2^N$ configurations of oriented loops, and then,
for each edge of the lattice, assigning height differences $\Delta
h=0,\pm\pi$ between the neighbouring sites of the dual lattice
according to whether the edge is contained in the oriented
subgraph $\cal G$, and its orientation. The weight $n$ (or $\sqrt
Q$) for each non-oriented loop is distributed into a factor
$e^{\pm i\chi}$ for each clockwise (anticlockwise) oriented loop,
where $n=2\cos\chi$. Although these weights are complex (a feature
which lies at the heart of the difficulties associated with a
rigorous treatment of the Coulomb gas approach), they have the
advantage of being local, in the sense that they may be
distributed so that each loop acquires a factor
$e^{i\theta\chi/2\pi}$ whenever it turns through an angle $\theta$
at a vertex.

However, it should be noted that, at least for the fully packed
model on the square lattice, there is a mapping to the 6-vertex
model with positive Boltzmann weights: at each vertex the two
loops are either oriented parallel to each other, with weight
$e^{i\chi/4}\cdot e^{-i\chi/4}+ e^{-i\chi/4}\cdot e^{i\chi/4}=2$,
or anti-parallel in which case the weight is
$$
e^{i\chi/4}\cdot e^{i\chi/4}+ e^{-i\chi/4}\cdot
e^{-i\chi/4}=2\cos(\chi/2)=(n+2)^{1/2}\,.
$$

The Coulomb gas method assumes that, in the continuum limit, the
discrete heights become continuous and the Boltzmann weights
converge to $e^{-S}$ where $S$ is the action of a free field
theory
$$
S=(g/4\pi)\int(\partial h)^2\,dxdy\,.
$$
The original discrete model may be recovered from this by adding a
term $\lambda\sum_R\cos2h(R)$ in the limit $\lambda\to-\infty$.

 However, on a cylinder of length $\ell$ and circumference $L$,
with $\ell\gg L$, this does not properly account for loops which
wind around it: these can be taken into account by placing
`electric' charges $e^{\pm i(\chi/\pi)h}$ at either end. This
modifies the partition function to $Z\sim e^{\pi c\ell/6L}$,
identifying the total central charge
$$
c=1-6\frac{(\chi/\pi)^2}g\,.
$$
The scaling dimensions of electric charges $e^{iqh}$ are also
modified if we put them at the ends of the cylinder as well:
$$
x_q=(1/2g)\left((q+\chi/\pi)^2-(\chi/\pi)^2\right)\,.
$$
Note that $x_q\not=x_{-q}$: this an example of the electric charge
asymmetry introduced by this construction.

$g$ is fixed in terms of $\chi$ by requiring \cite{Kondev} that
$\cos2h$ be marginal in the sense of the renormalisation group,
i.e.~$x_2=2$. This fixes
$$
g=1\pm(\chi/\pi)\,,
$$
with the sign depending on whether we choose $x_2$ or $x_{-2}$. In
fact this ambiguity is to be expected: for each value of $n$,
$\chi$ is only defined up to a sign (actually we can add multiples
of $2\pi$ as well, but these give less relevant operators) and
these correspond to the dilute ($g>1$) and dense ($g<1$) cases of
the critical O$(n)$ model. In the following we take the lower sign
by convention.

Note that these ideas are easy to extend to the partition function
on the torus, correctly taking into account loops which wrap
around some combination of the two cycles \cite{diFSZ}.

Now consider the case of the annulus. Throughout this paper we
assume free boundary conditions on the O$(n)$ spins, which means
that there are only closed loops in the loop gas representation.
(Partial results using Coulomb gas methods were found for the case
of fixed boundary conditions in Ref.~\cite{JCcrossing} for the
special case $n=1$ in the dense phase.)

First consider the case when $\ell\gg L$, where we expect
$$
Z\sim e^{\pi c\ell/24L}\sim q^{-c/24}\,,
$$
with $c$ given as above. In this limit there is no contribution of
loops wrapping around the annulus, so we expect that that
$h(y=L)=h(y=0)$. Naively then, we get a free field theory with
(equal) Dirichlet boundary conditions, which gives $c=1$.

Where does the correction to $c$ come from? One (incorrect)
possibility is as follows: looking back at the lattice
construction, we see that there are extra factors of $e^{\pm
i\chi/2}$ whenever a loop is next to the boundary, which are not
properly taken into account in the the bulk Boltzmann weights of
the height model. The sign is determined by whether the height at
a site next to the boundary is $\pm\pi$. In the continuum limit
these would lead to boundary terms in the action proportional to
$i\chi\int \partial_\perp h\, dl$ where $\partial_\perp$ is along
the inward pointing normal to the boundary and $dl$ is a line
element. For the annulus these would give something proportional
to
\begin{equation}
\label{b1} i\chi\int(\partial_yh(x,y=0)-\partial_yh(x,y=L))dx
\end{equation}
However, an explicit calculation (see Appendix) shows that such a
combination does not contribute to $c$. In fact, if we add to the
action a general boundary term
$$
\int(\alpha_1\partial_yh(x,y=0)+\alpha_2\partial_yh(x,y=L))dx\,,
$$
we find that the effective central charge is
$$
c=1-(24/g)(\alpha_1+\alpha_2)^2\,.
$$
Thus not only is there no contribution if $\alpha_1=-\alpha_2$, as
in (\ref{b1}), we must also have $\alpha_1+\alpha_2$ real, rather
than pure imaginary.

An equivalent, and easier, way of getting the same modification to
$c$ is to assume that the correct boundary conditions, even when
$\ell\gg L$, are
$$
h(y=L)-h(y=0)=\pi m_0\not=0\,.
$$
In that case we can write $h=\pi m_0y/L+\tilde h$, where $\tilde
h$ vanishes on both $y=0$ and $y=L$. The functional integral over
$\tilde h$ gives $c=1$ as before, and the modification to the
partition function is $\sim\exp(-(g/4\pi)(\pi m_0)^2\ell/L)$. So
if we take
\begin{equation}
\label{m0} m_0=\pm\chi/\pi g\,,
\end{equation}
we get the correct $c$. Note that, for a long rectangular strip
rather than an annulus, this is like adding magnetic charges, or
vortices, at the ends.

Thus it would appear that there should be a spontaneous average
magnetic flux around the annulus, if $g\not=1$. This may be
understood heuristically in terms of the preferred parallel,
rather than antiparallel, alignment of neighbouring loops: this
effect should be enhanced near a boundary, since loops are
geometrically constrained to lie approximately parallel or
anti-parallel to the boundary.

\begin{figure}[t]
\centering
\includegraphics[width=10cm]{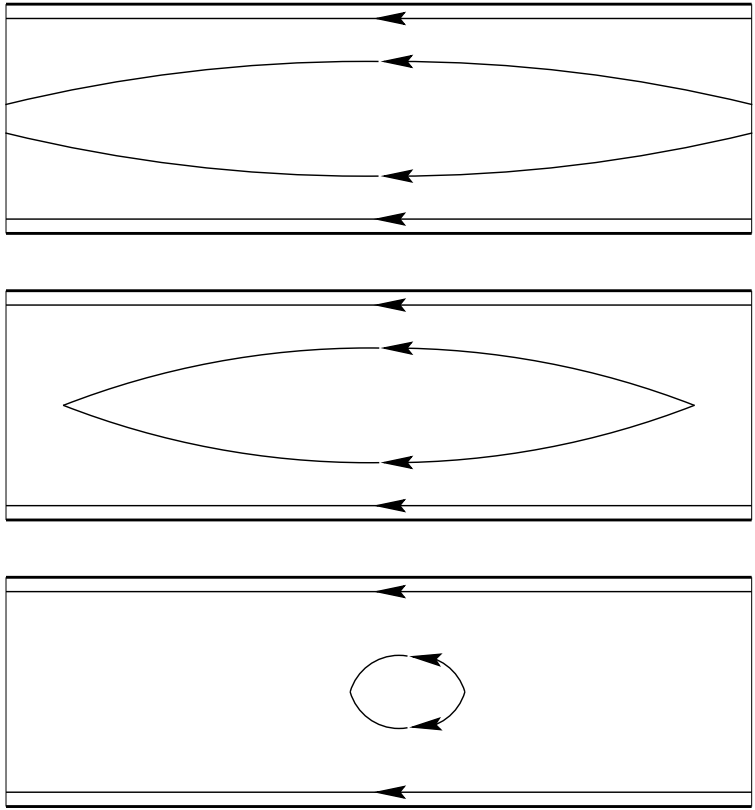}
\caption{\label{screening}\small The screening of magnetic flux in
a long rectangle. If the background flux $m$ is too large, vortex
pairs of strength $\pm2$ can be shed from either end of the
rectangle and will annihilate in order to reduce the free energy.
If $m$ is too small, the opposite effect occurs.}
\end{figure}

The actual value of $m_0$ may be fixed by the following argument.
Consider first the geometry of a long rectangle with $\ell\gg L$,
where the loops are allowed to end on the boundaries at
$x=0,\ell$, but, as before, not on $y=0,L$. In that case the total
charge $m$ flowing along the rectangle is not fixed, and we can
ask the question what is its mean value $m_0$ in the state of
lowest free energy. A flux $m$ corresponds to vortices of
strengths $\pm m$ at the ends of the rectangle. However this can
increase or decrease in units of 2 by shedding vortices from
either end of the rectangle (see Fig.~\ref{screening}). The
additional free energy for creating such a pair of vortices is
$(g/4\pi)\big((m+2)^2-m^2\big)(\pi/L)^2(\ell L)$ scaling dimension
of a vortex of strength $\pm2$, from which we read off the scaling
dimension
$$
\Delta_2=(g/4)\big((m+2)^2-m^2\big)
$$
If $m$ is too large, $\Delta_2>1$, which means that the
corresponding renormalisation group $y_2=1-\Delta_2<0$. (Note that
since we are in the boundary, rather than the bulk, situation, the
eigenvalue is $1-\Delta$ rather than $2-\Delta$.) This implies
that such $\pm2$ vortex pairs are closely bound. Thus any such
pairs shed from the boundaries at $x=0,\ell$ will annihilate to
reduce their free energy. On the other hand, if $m$ is too small,
$y_2>0$, which means that any such vortex pairs will unbind. This
will act to increase the effective value of $m$.

This screening effect implies that the mean value $m_0$ of the
magnetic flux which minimises the free energy corresponds to
$y_2=0$, that is
$$
(g/4)((m_0+2)^2-m_0^2)=1\,,
$$
so that
$$
m_0= (1-g)/g=\chi/\pi g\,,
$$
which is the same as found above in (\ref{m0}). There is a similar
minimum free energy solution with $m=-m_0$. Note that the above
argument is analogous to the earlier argument which fixed $g$,
where we demanded that electric $\pm2$ charges should be marginal.

If we now go the annulus, we expect a total average magnetic flux
$\pm m_0$ to spontaneously form, even when $\ell\gg L$. Now
suppose that $\ell/L$ is not so large, so we can have extra loops
wrapping around the $x$-cycle. We can orient these as before. If
the total number of up arrows minus down arrows (the additional
magnetic flux flowing along the annulus) is $p$, then we get
$h(y=L)-h(y=0)=\pi (p\pm m_0)$. As for the cylinder, in order to
count them correctly we need to put in a factor $\exp(i(p\pm
m_0)\chi)$. Thus we get the following first guess for the
partition function on the annulus:
$$
\widetilde Z=Z_0\sum_{p\in\mathbb Z}e^{i(p+m_0)\chi}e^{-(g/4)(\pi
p+m_0)^2(\pi\ell/L)}+(m_0\to -m_0)\,,
$$
where $Z_0=q^{-1/24}\prod_{r=1}^\infty(1-q^r)^{-1}$ is the
partition function from $\tilde h$. Note that we should sum over
both possible signs for $m_0$. If we let $p\to-p$ in the first
term this simplifies to
\begin{equation}
\label{1} \widetilde
Z=q^{-c/24}\prod_{r=1}^\infty(1-q^r)^{-1}\sum_{p\in\mathbb
Z}\cos((p-m_0)\chi)\, q^{(g/4)p^2-(1-g)p/2}\,.
\end{equation}
Note that if we want to count loops wrapping around the annulus
with a different weight $n'=2\cos\chi'$, we just change
$\chi\to\chi'$ in the above (keeping $g$ the same.)

Eq.~(\ref{1}) has some good features and some bad ones. In general
we expect that $Z$ can be written as a sum of terms $q^h$ where
$h$ runs over all the allowed scaling dimensions of the allowed
boundary operators. (For a unitary theory the coefficients should
be non-negative integers, but this doesn't have to hold for
general $n$.) We see in (\ref{1}) for $p=N\geq1$ the scaling
dimensions of the boundary $N$-leg operators, as first conjectured
by Saleur and Duplantier \cite{DupSal}.

However, for the dilute case with $g>1$, $p=-1$ actually gives the
next-to-leading term as $q\to0$. This doesn't make sense: we
expect this to come from $p=N=1$. More seriously, (\ref{1}) fails
to account for the fact that the scaling dimensions of the
boundary $N$-leg operators correspond to those of the degenerate
cases $h_{1,N+1}$ of the Kac table: in general these operators
correspond to highest weight states whose Virasoro representations
are reducible: they have a null descendent state (which
corresponds to a term in the expansion of $\prod_r(1-q^r)^{-1}$)
of dimension $h_{1,N+1}+N+1$. For a unitary theory, or more
generally a minimal model, we know that such states (and all their
descendents) should be subtracted out of the partition function.
For a non-unitary theory this is not necessary: however we shall
show later that if they are retained, the behaviour as $\tilde
q\to0$ is incorrect. This leads to the conclusion that each term
in (\ref{1}) should be modified according to
$$
q^{(g/4)p^2-(1-g)p/2}\to
q^{(g/4)p^2-(1-g)p/2}\,\big(1-q^{p+1}\big)=
q^{h_{1,p+1}}-q^{h_{1,-p-1}}\,.
$$
Note that this has the feature of automatically eliminating the
`rogue' state at $p=-1$.

\begin{figure}[t]
\centering
\includegraphics[width=10cm]{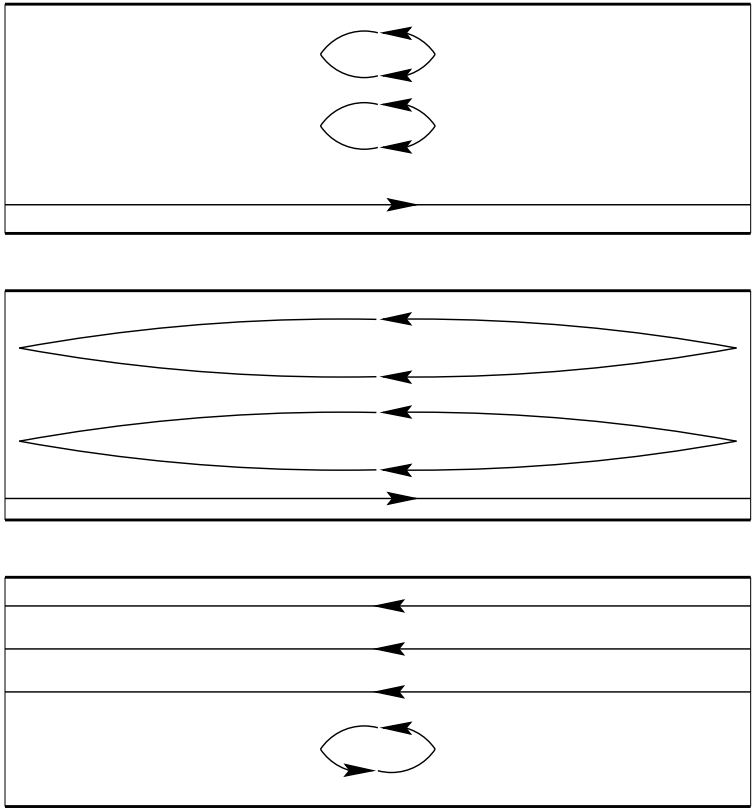}
\caption{\label{null}\small Mechanism for the appearance of null
states. In this example, an excited state in the $p=1$ sector has
just sufficient energy $2\pi/L$ to allow two marginally bound
pairs of $\pm2$ vortices to form. These can then move to the ends
of the rectangle, and one of them can then annihilate with the
original flux line. The state is therefore equivalent to the
ground state in the $p=-3$ sector, which has already been counted
in (\ref{1}) and which therefore must be subtracted off.}
\end{figure}

We now give a physical argument for this subtraction. Once again
it is useful to think about the rectangle geometry where magnetic
flux can be created or destroyed at the boundaries at $x=0,\ell$.
It is also useful to think in terms of the energy eigenstates of
the hamiltonian $(\pi/L)(L_0-c/24)$ which generates translations
in $x$. In general, each configuration with total magnetic flux
$p$ will be accompanied by excitations of the $\tilde h$ field,
which correspond to the Virasoro descendents. If the energy of
these is correct they can resonate with the original highest
weight state plus a number of pairs of marginally bound $\pm 2$
magnetic charges, each of which has energy $\pi/L$. If the
excitation energy is $(p+1)\pi/L$, exactly $p+1$ vortex pairs can
be shed from the boundaries at $x=0,\ell$ (see Fig.~\ref{null}).
These states, however, are identical to those with total magnetic
charge $p-2(p+1)=-2-p$, and should therefore not be doubly
counted.

The effect of this subtraction is therefore to modify the sum in
(\ref{2}) to
$$
\sum_{p\in\mathbb
Z}\cos((p-m_0)\chi)\,\big(q^{h_{1,p+1}}-q^{h_{1,-p-1}}\big)\,.
$$
Relabelling $p\to-2-p$ in the second term has the effect of
modifying
$$
\cos((p-m_0)\chi)\to\cos((p-m_0)\chi)-\cos((p+2+m_0))\chi)\propto\sin((p+1)\chi)\,,
$$
which finally leads to the conjecture (\ref{2}) for the annulus
partition function, after normalising so that the coefficient of
the $p=0$ term (the contribution of the identity operator) is
unity:
$$
Z=q^{-\frac c{24}}\prod_{r=1}^\infty(1-q^r)^{-1} \sum_{p\in\mathbb
Z}\frac{\sin(p+1)\chi}{\sin\chi} \, q^{\frac
{gp^2}4-\frac{(1-g)p}2}\,.
$$

Note that all the coefficients in (\ref{2}) are polynomials in
$n=2\cos\chi$ as we expect. For $p=1$ we get exactly $n$ (the
degeneracy factor for a single loop wrapping around the annulus),
for $p=2$ we get $n^2-1$, and so on.

\subsection{Modular properties}
Now we express (\ref{2}) (for general $\chi'$) in terms of the
conjugate modulus $\tilde q=e^{-2\pi L/\ell}$. Setting
$q=e^{-\delta}$ we have
$$
Z=Z_0\,q^{\frac{1-c}{24}}(\sin\chi')^{-1}{\rm Im}\,e^{i\chi'}
\sum_pe^{ip\chi'}e^{-\delta(\frac{gp^2}4-\frac{(1-g)p}2)}\,.
$$
Using the Poisson sum formula $\sum_p\to\sum_m\int dp\,e^{2\pi
imp}$, the integral is
\begin{eqnarray*}
&&\int e^{-\frac{\delta gp^2}4+[\frac{(1-g)\delta}2+i\chi'+2\pi im]p}dp\\
&&=(4\pi/\delta g)^{1/2}e^{-\frac 1{\delta g}[\chi'+2\pi
m-i\frac{(1-g)\delta}2]^2}\\
&&=(4\pi/\delta g)^{1/2}e^{-\frac1{\delta g}(\chi'+2\pi m)^2
+i(\chi'+2\pi m)\frac{(1-g)}g +\frac{(1-g)^2\delta}{2g}}\,.
\end{eqnarray*}
The last term in the exponential cancels the $q^{(1-c)/24}$. Under
a modular transformation
$$
Z_0=(\delta/2\pi)^{1/2}{\tilde q}^{-\frac1{12}}\prod_r(1-{\tilde
q}^{2r})^{-1}\,,
$$
so we end up with
$$
 Z=(2/g)^{1/2}{\tilde q}^{-\frac
c{12}}\prod_r(1-{\tilde q}^{2r})^{-1}
\sum_m\frac{\sin((2m\chi+\chi')/g)}{\sin\chi'} {\tilde
q}^{\frac{(\chi'+2\pi m)^2-\chi^2}{2\pi^2g}}\,,
$$
which finally simplifies to (\ref{mod}). Note that if we had not
subtracted off the null states, as in (\ref{1}), we would find
$\widetilde Z\sim{\tilde q}^{-1/12}$ rather than ${\tilde
q}^{-c/12}$. This is an example of how the null state subtraction
is necessary to maintain the correct modular properties
\cite{JC86}.

The leading term as $\tilde q\to0$, with $m=0$, agrees with the
exponent found in Ref.~\cite{JClink} for the case of loops
wrapping around a long cylinder counted with weight
$n'=2\cos\chi'$. We now have also the prefactor:
\begin{equation}
\label{3} Z\sim (2/g)^{1/2}\frac{\sin(\chi'/g)}{\sin\chi'} {\tilde
q}^{\frac{{\chi'}^2-\chi^2}{2\pi^2g}}\,.
\end{equation}

If we set $\chi'=\chi$, the other exponents in (\ref{mod}) are
those of \em even \em electric charge operators $x_{2m}$. For
general $g$ these are not in the Kac table, consistent with the
fact that there is no explicit substraction of null states in
(\ref{mod}), so the characters are simply given by the infinite
product.

We now check (\ref{2}) against some known cases.

\section{Comparison with known results.}
\label{sec3}
\subsection{$n=0$ in the dilute regime.}
In this  case $g=3/2$, $\chi=-\pi/2$, for which we expect $Z=1$,
since loops wrapping round the annulus should all get a weight
$n=0$. (\ref{2}) gives
$$
Z=\prod_r(1-q^r)^{-1}\sum_p\sin((p+1)\pi/2)\,q^{\frac{3p^2}{8}+\frac
p4}\,.
$$
The prefactor is $+1$ if $p\equiv0\pmod{4}$, $-1$ if
$p\equiv2\pmod{4}$, and zero otherwise. After a little algebra we
get
$$
Z=\frac{\sum_{k\in\mathbb
Z}(q^{6k^2+k}-q^{6k^2+5mk+1})}{\prod_r(1-q^r)}\,.
$$
This is identically equal to 1 by Euler's pentagonal identity.

\subsection{$n=1$, dilute phase.}
This should correspond to the unitary CFT which describes the
scaling limit of the critical Ising model. Now $\chi=-\pi/3$,
$g=4/3$. The numerator in (\ref{2}) is
$$
\sum_p\frac{\sin\frac{(p+1)\pi}3}{\sin\frac\pi
3}\,q^{\frac{p^2}3+\frac p6}\,.
$$
Now
\begin{eqnarray*}
p=6k&&\mbox{gives\ }\, q^{12k^2+k}\\
p=6k-2&:& -q^{12k^2-7k+1}\\
p=6k+1&:& q^{12k^2+5k+\frac12}\\
p=6k+3&:& -q^{12k^2+13k+\frac72}\,.
\end{eqnarray*}
Using the Rocha-Caridi character formula \cite{BYB}
$$
\chi_{r,s}(q)=\prod_r(1-q^r)^{-1} \sum_{k\in\mathbb Z}
\left(q^{\frac{(24k+4r-3s)^2-1}{48}}- \{s\to -s\}\right)\,,
$$
for the case $c=\frac12$, we see that the first 2 terms give
$\chi_{1,1}$ and the second pair give $\chi_{1,3}$. This then
agrees with the result \cite{JC89} for the Ising model with free
boundary conditions
$$
Z=\chi_{1,1}+\chi_{1,3}\,.
$$
Alternatively we can look at the dual spins, which are fixed on
the boundary. If they are fixed into the same state on both
boundaries we must have $p$ even, so that $Z=\chi_{1,1}$, and if
they are fixed into opposite states $p$ must be odd, so
$Z=\chi_{1,3}$. These also agree with Ref.~\cite{JC89}.

\subsection{$n=2$}
In this case $\chi=0$ and $g=1$, so $\sin(p+1)\chi/\sin\chi\to
p+1$. The numerator in (\ref{2}) becomes
$$
\sum_p(p+1)q^{\frac{p^2}4}=\sum_{p\in\mathbb Z} q^{\frac{p^2}4}\,.
$$
This agrees with the interpretation as the XY model at the
Kosterlitz-Thouless transition: the terms with $p\not=0$
correspond to a total vorticity $\pm p$ along the annulus.

\subsection{$Q=3$ Potts model.}
Now $\chi=\pi/6$ and $g=\frac56$. The numerator in (\ref{2}) is
$$
\sum_p\frac{\sin((p+1)\pi/6)}{\sin\pi/6}\,q^{\frac{5p^2}{24}-\frac
p{12}}\,.
$$
If we take free boundary conditions on both boundaries we should
restrict $p$ to be even. Then
\begin{eqnarray*}
p=12k&:&q^{30k^2-k}\\
p=12k+2&:&2q^{30k^2+9k+\frac23}\\
p=12k+4&:&q^{30k^2+19k+3}\\
p=12k+6&:&-q^{30k^2+29k+7}\\
p=12k-4&:&-2q^{30k^2-21k+frac{11}3}\\
p=12k-2&:&-q^{30k^2-11k+1}\,.
\end{eqnarray*}
These pair up as follows: $((1,6),(2,5),(3,4))$ to give
$$
Z=\chi_{1,1}+2\chi_{1,3}+\chi_{1,5}\,,
$$
which agrees with Ref.~\cite{JC89}.

Note that if we choose free boundary conditions on one edge and
fixed on the other, $p$ is restricted to be odd, and the leading
term as $q\to0$ comes from $p=1$, and is
$$
Z\sim\sqrt3\,q^{\frac18}\,.
$$
The $\sqrt3$ is to be expected, because in the Fortuin-Kasteleyn
representation each closed loop carries a factor $\sqrt Q$.

\subsection*{$n=Q=1$, dense phase}
In this case $\chi=\pi/3$, $g=2/3$. The numerator in (\ref{2}) is
$$
\sum_p\frac{\sin((p+1)\pi/3)}{\sin(\pi/3)}\,q^{\frac{p^2}6-\frac
p6}\,.
$$
We get a non-zero contribution in the following cases:
\begin{eqnarray*}
p=6r&:& q^{6r^2-r}\\
p=6r-2&:& -q^{6r^2-5r+1}\\
p=6r+1&:& q^{6r^2+r}\\
p=6r+3&:& -q^{6r^2+5r+1}\,.
\end{eqnarray*}
Using Euler's identity again we see that $Z=2$, consistent with
the dual interpretation as the Ising model at zero temperature.
(The factor 2 is due to the global spin reversal.) On the other
hand this model can be interpreted as the $Q=1$ Potts model
(percolation). Choosing the sites on both boundaries to be in the
same Potts state enforces $p$ to be even, and then we get $Z=1$ as
expected.

\subsection*{$n=Q=0$, dense phase}
Now $\chi=\pi/2$, $g=\frac12$. The numerator in (\ref{2}) is
$$
\sum_p\sin(p+1)\pi/2\, q^{\frac{p^2}8-\frac p4}\,,
$$
so $p$ is even. For
\begin{eqnarray*}
p=4k&:&q^{2k^2-k}\\
p=4k+2&:&-q^{2k^2+k}\,,
\end{eqnarray*}
so
$$
Z=\sum_k(q^{2k^2-k}-q^{2k^2+k})=0\,.
$$
This is correct, since in this case there is just one macroscopic
loop (or spanning tree) which is counted with weight $n=0$.

\section{Some new results.}
\label{sec4}
\subsection{Percolation.}
By setting $\cos\chi=0$ in (\ref{2}) with $g=\frac23$ we suppress
all other contributions with a non-zero number of loops wrapping
around the annulus. In terms of percolation, this happens if and
only if there exists a cluster connecting the two boundaries. This
crossing probability is therefore
\begin{eqnarray*}
P&=&\prod_{r=1}^\infty(1-q^r)^{-1}\sum_p\sin((p+1)\pi/2)\,q^{\frac{p^2}6-\frac
p6}\\
&=&\prod_{r=1}^\infty(1-q^r)^{-1}\sum_{k\in\mathbb
Z}\big(q^{\frac{8k^2}3-\frac{2k}3}
-q^{\frac{8k^2}3+2k+\frac13}\big)\,,
\end{eqnarray*}
so that $1-P\sim q^{1/3}$ as $q\to0$. Using the Jacobi triple
product formula this can be written in terms of the Dedekind
function $\eta(\tau)\equiv q^{1/24}\prod_{r=1}^\infty(1-q^r)$ with
$q=e^{-2\pi i/\tau}$ as
$$
P=\frac{\eta(-1/3\tau)\eta(-4/3\tau)}{\eta(-1/\tau)\eta(-2/3\eta)}
=(3/2)^{1/2}\,\frac{\eta(3\tau)\eta(3\tau/4)}{\eta(\tau)\eta(3\eta/2)}\,.
$$

In the opposite limit, using (\ref{4}) or the above, we have
$$
P\sim (3/2)^{1/2}\,{\tilde q}^{\frac5{48}}\,,
$$
as $\tilde q\to0$, which is the well-known `magnetic' exponent
\cite{denNijs} for the $Q=1$ Potts model (also known as the 1-arm
exponent \cite{1arm} in the SLE literature.) Note that this result
is different from, and much larger than, the result found in
Ref.~\cite{JCcrossing}. This is because in that paper crossing
clusters which also wrap around the annulus were disallowed. It
would be interesting to compare the above result with the implicit
formula derived by Dub\'edat \cite{Dub} using SLE methods.

Note that, in principle, one can solve for $e^{i\chi'}$ as a
function of $n'$ and substitute in (\ref{2}), hence obtaining the
complete generating function for the probabilities that a given
number of clusters wrap around the annulus.

\subsection{Self-avoiding loop: dilute case.}
If we take the $O(n')$ term in (\ref{2}) with $g=\frac32$ we
obtain the partition function $Z_1$ for a single self-avoiding
loop which wraps around the annulus. From (\ref{2}) we need
$$
\frac\partial{\partial
n'}\left.\frac{\sin((p+1)\chi')}{\sin\chi'}\right|_{\chi'=-\pi/2}
=-\frac12(p+1)\cos((p+1)\pi/2)\,.
$$
So we get a non-zero result only when $p$ is odd, say $p=2k-1$,
whence, after a little algebra,
\begin{equation}
\label{4} Z_1=\prod_{r=1}^\infty(1-q^r)^{-1}\sum_k{k\in\mathbb
Z}k(-1)^{k-1}\,q^{\frac{3k^2}2-k+\frac18}\,.
\end{equation}
The leading behaviour as $q\to0$ comes from $k=1$ and is
$$
Z_1\sim q^{\frac58}\,,
$$
as expected.

In the opposite limit we can use (\ref{3}). In this case the
leading behaviour comes from differentiating the exponent:
$$
Z_1\sim\frac12(2/g)^{1/2}\frac{\sin(\chi/g)}{\sin\chi}\frac{\chi}{g\pi^2}\ln\tilde
q=\frac1{6\pi} |\ln\tilde q|\,.
$$
If the annulus is mapped into the region between two circles radii
$r_1$ and $r_2>r_1$, the last factor is just $\ln(r_2/r_1)$.

\subsection{Self-avoiding loop: dense phase.}
In this case
$$
Z_1=-q^{-c/24}\prod(1-q^r)^{-1}\sum_p\frac{p+1}2\cos((p+1)\pi/2)\,
q^{\frac{p^2}8-\frac p4}\,.
$$
If we let $p\to2-p$ we get the same expression except
$(p+1)\to(3-p)$. So the sum is
$$
-\sum_p\cos((p+1)\pi/2)\,q^{\frac{p^2}8-\frac p4}\,,
$$
and finally (since $c=-2$)
$$
Z_1=q^{1/12}\prod_r(1-q^r)^{-1}\sum_k\big(q^{2k^2-\frac18}-q^{2k^2-2k+\frac38}\big)\,.
$$
Using the Jacobi triple product formula this can be rewritten as
$$
Z_1=q^{-\frac1{24}}\prod_{m=1}^\infty(1-q^{m-\frac12})^2\,.
$$
The leading term as $q\to0$ is
$$
Z_1\sim q^{-\frac1{24}}\,.
$$
This is reasonable since the loop is weighted by a factor
$\mu^{-{\rm length}}$ so the contribution grows exponentially with
$\ell$.

\subsection{Logarithmic cases.}
Logarithmic CFTs have been studied for some time, although the
question of how they satisfy modular invariance has not been
resolved in general \cite{Flohr}. The limit $n\to0$ of the O$(n)$
model affords an example in both the dilute and dense regimes;
other examples have been discussed in Ref.~\cite{ReadSal}. If we
differentiate the whole expression for $Z$ wrt $n$ at $n=0$, we
get 3 kinds of contribution: the first comes from differentiating
wrt $\chi'$: this gives the partition function for loops wrapping
round the annulus as found above, and is a regular series in $q$.
The second comes from differentiating $q^{-c/24}$ and gives
$-(c'(0)/24)\ln q$ times the usual partition function at $n=0$.
The third comes from differentiating the exponents: using
$$
\frac\partial{\partial
g}(\frac{gp^2}4+\frac{(g-1)p}2)=\frac{p^2}4+\frac p2\,,
$$
the result, in the dilute case when $g=\frac32$, is proportional
to
$$
(\ln q)\sum_p(p^2+2p)\sin((p+1)\pi/2)\,q^{\frac{3p^2}8+\frac
p4}\,.
$$
As before, the only contributions come from $p$ even, and
proceeding as before we get
$$
\ln q\,\sum_k(k(2k+1)q^{6k^2+k}-k(2k-1)q^{6k^2-5k+1})\,,
$$
or
$$
Z_{\rm log}\propto\ln
q\,\sum_kk(2k+1)(q^{6k^2+k}-q^{6k^2+5k+1})\,.
$$
Note that we still get the null state structure in this
logarithmic sector.

In the dense phase, the partition function vanishes. The
non-logarithmic terms have already been evaluated. The logarithmic
term is very similar to the above:
$$
\ln q\,\sum_k(k(2k+1)q^{2k^2+k}-k(2k-1)q^{2k^2-3k+1})\,,
$$
or
$$
Z_{\rm log}\propto\ln q\,
\sum_kk(2k+1)(q^{2k^2+k}-q^{2k^2+3k+1})\,.
$$
once again showing the null states. The leading term as $q\to0$
gives the contribution of single dense loops (or spanning trees)
which do not wrap around the annulus: note that this is much
smaller than the $O(q^{-1/24})$ contribution from those which do.

\section{Summary and further remarks.}
In this paper we have presented an explicit result for scaling
limit of the partition function of the critical O$(n)$ and
$Q$-state Potts models on the annulus. Our formalism makes it
simple to count loops which wrap around the annulus with different
weights, leading potentially to many new formula for crossing
probabilities in percolation and for self-avoiding loops, some of
which have been presented here.

The electric-magnetic dual of our arguments for a long ($\ell\gg
L$) annulus can in fact be applied to a long ($L\gg\ell$)
cylinder, to give an alternative derivation of the usual relation
between $g$, $\chi$ and $n$. For a long annulus we argued that
$\pm2$ magnetic charges (vortices) would rearrange themselves in
such a way as to induce mean magnetic charges $\pm m_0=\pm(1-g)/g$
at the ends of a long rectangle, leading to a net magnetic flux
around the annulus. For a long cylinder, one may similarly argue
that the $\pm2$ electric charges in the model also rearrange
themselves to give net electric charges $\pm e_0=\pm(1-g)$ at
either end. These can then be interpreted as counting loops going
around the cylinder with the weight $n=2\cos\pi e_0$.

Like all Coulomb gas methods, however, the arguments are somewhat
heuristic, although they lead to completely explicit formulae,
and, because of the complex weights, it seems hard to make them
rigorous. In particular, although we have computed the partition
function, it is by no means clear that the same ensemble can be
used to compute correlation functions in the original model. It
would be nice to see a direct connection with the other `Coulomb
gas' approach which has been employed in CFT, namely that
originally developed by Dotsenko and Fateev \cite{DotFat}. This is
essentially a way of constructing holomorphic conformal blocks
using modified free field theory. However, it has many features in
common with the Coulomb gas construction used here: the background
charge $m_0$, and the marginal screening operators, which in our
case are the $\pm2$ vortices. In the bulk, it is still necessary,
in the Dotsenko-Fateev approach, to sew together the holomorphic
and anti-holomorphic blocks in a consistent way to obtain
correlation functions, but in boundary CFT the correlation
functions are linear combinations of the conformal blocks
(specialised to real values of their arguments), and so the
correpondence between the two approaches should be more direct. It
would, of course, be important to establish any of these results
rigorously, for example by using SLE methods.

\noindent\em Acknowledgement\em. This work was supported in part
by EPSRC Grant GR/R83712/01. I thank H.~Saleur for pointing out
several important references missing in an earlier version of this
paper.

\appendix
\section{Boundary terms in the gaussian model.}
\label{app} Consider the action
$$
S=S_0+S_1=\frac g{4\pi}\int(\nabla
h)^2dxdy+\alpha_1\int\partial_yh(x,y=0)dx
+\alpha_2\int\partial_yh(x,y=L)dx\,.
$$
We wish to compute the regularised free energy, or equivalently
the ground state energy $E_0$ of the associated hamiltonian. Let
$$
h(x,y)=\sum_{n=1}^\infty\frac{f_n(x)}{(gL/4\pi)^{1/2}}\,\sin\frac{n\pi
y}L\,.
$$
Then
$$
S_0=\int dx\sum_n(\frac12{\dot f}_n^2+\frac12(n\pi/L)^2f_n^2)\,,
$$
from which we read off the ground state energy
$$
E_0=\frac12\sum_n\omega(n\pi/L)\,,
$$
where $\omega(k)=k$. We can regularise this sum either by
modifying the dispersion relation (eg using a lattice, in which
case $\omega(k)=2\sin(k/2)$) and using the Euler-Maclaurin
formula, or using zeta-function, in which case we get the standard
result
$$
E_0=\frac\pi{2L}\zeta(-1)=-\frac\pi{24L}\,,
$$
corresponding to $c=1$.

Now add in
$$
S_1=(4\pi/gL)^{1/2}\sum_n(n\pi/L)(\alpha_1+(-1)^n\alpha_2)f_n\,.
$$
Completing the square, the contribution of the $n$th mode can be
written
$$
\frac12\left((n\pi/L)f_n+(4\pi/gL)^{1/2}(\alpha_1+(-1)^n\alpha_2)\right)^2
-\frac{2\pi}{gL}(\alpha_1+(-1)^n\alpha_2)^2\,,
$$
so the change in the ground state energy is
$$
E_1=-(2\pi/gL)\left(\sum_{n {\rm odd}}(\alpha_1-\alpha_2)^2
+\sum_{n {\rm even}}(\alpha_1+\alpha_2)^2\right)\,.
$$

If we use zeta function regularisation we have
\begin{eqnarray*}
\sum_{n {\rm odd}}1&=&\lim_{s\to0}(\sum_nn^{-s}-\sum_n(2n)^{-s})
=\lim_{s\to0}(1-2^{-s})\zeta(s)=0\\
\sum_{n {\rm
even}}1&=&\lim_{s\to0}\sum_n(2n)^{-s}=\zeta(0)=-\frac12\,,
\end{eqnarray*}
so
\begin{equation}
\label{E1} E_1=\frac\pi{gL}(\alpha_1+\alpha_2)^2\,.
\end{equation}
If we use a lattice dispersion relation and also replace the
boundary derivatives by finite differences, the sum becomes
$$
\sum_{n=1}^{L-1}\frac{(\alpha_1+(-1)^n\alpha_2)^2}{\cos^2(n\pi/2L)}\,,
$$
on which we use the formulae
\begin{eqnarray*}
\sum_{n {\rm odd}}f(n/L)&=&\frac12L\int_0^1f(x)dx+O(L^{-2})\,,\\
\sum_{n {\rm even}}f(n/L)&=&\frac12L\int_0^1f(x)dx-\frac1{2L}f(0)+
O(L^{-2})\,,
\end{eqnarray*}
giving the same result. It can also be verified by writing, for a
general position dependent $\alpha$
$$
Z=Z_0\left\langle\exp\left(\int\alpha(l)\partial_\perp
h(l)dl\right)\right\rangle =Z_0
\exp\left(\frac12\int\int\alpha(l)\alpha(l')\partial_\perp
\partial'_\perp G(l,l')dldl'\right)\,,
$$
where $G$ is the Green's function for the free field with
Dirichlet boundary conditions.

(\ref{E1}) leads to a modification to the effective central charge
$$
c=1-\frac{24}g(\alpha_1+\alpha_2)^2\,,
$$
as claimed in Sec.~\ref{sec2}.

\end{document}